\documentstyle[12pt]{article}
\begin{document}
\begin{flushright}          
AUE-00/02     
\end{flushright}     
\vskip 0.3cm     
\centerline{\Large\bf
Zee-type Neutrino Mass Matrix \\
and\\
 Bi-Maximal Mixing}
\vskip 2cm
\begin{center}
{\bf Masahisa Matsuda\footnote{E-mail address:mmatsuda@auecc.aichi-edu.ac.jp}}\\
{\it Department of Physics and Astronomy, Aichi University of Education, 448-8542 
Kariya, Japan}\\
{\bf Cecilia Jarlskog\footnote{E-mail address:ceja@matfys.lth.se} and Solveig Skadhauge
\footnote{E-mail address:solveig@matfys.lth.se}}\\
{\it Department of Mathematical Physics, LTH, Lund University, S-22100 Lund, Sweden}\\
{\bf Morimitsu Tanimoto{\footnote{E-mail address:tanimoto@muse.hep.sc.niigata-u.ac.jp}}\\
{\it Department of Physics, Niigata University, 950-2181 Niigta, Japan}}
\end{center}
\vskip 2 cm          
\centerline{\bf Abstract}\par
It is shown that the bi-maximal solution is the only possibility to reconcile 
Zee-type neutrino mass matrix with three flavors to the current atmospheric 
and solar neutrino experimental data.
The mass of the lightest neutrino, which consist mostly of $\nu_{\mu}$ and $\nu_{\tau}$, is $\simeq \Delta m_{\odot}^2/(2\sqrt{\Delta m_{atm}^2})$. 
The related topics on Zee-type neutrino mass matrix are also discussed.
\newpage
\section{Introduction}
Recent atmospheric\cite{Atm} and solar\cite{Sol} neutrino experiments suggest that
neutrinos have tiny masses compared to quarks and charged leptons and the MNS mixing matrix is 
different from the CKM matrix in the quark sector. 
The tiny masses for neutrinos are expected to be less than ${\cal O}(1)$eV. 
  It is well known that there are mainly two mechanisms to obtain tiny neutrino masses. 
One is the seesaw mechanism, which requires the existence of right-handed Majorana 
neutrinos with $10^{10}-10^{15}$GeV masses. The other mechanism is to generate tiny masses 
by higher order loop effects by the extension of the standard model(SM). 
In both senarios light neutrinos are Majorana. 
In this talk\cite{JMST}  we investigate the second possibility to interpret the tiny neutrino masses
and mixing patterns suggested by experiments\cite{Atm,Sol} within the 
framework of three flavors.
\par
\section{Zee Model}
A model with extension of the Higgs sector by adding one more doublet in $SU(2)_L$ 
and a charged singlet scalar with lepton number 2 was proposed by Zee\cite{Zee1} 
almost two decades ago. 
The Zee model is the minimal extension of the SM and the one loop effects by extended Higgs 
sector can induce small Majorana neutrino masses. 
In the Zee model the neutrino masses are  given by 
\begin{equation}
m_{ll'}=f_{ll'}(m_l^2-m_l'^2)\frac{\mu v_u}{v_d}F(M_1^2,M_2^2) ,
\label{Mnu}
\end{equation}
where $M_{1,2}$ are the masses of the two physical charged scalars. 
This type of neutrino matrix has the following 
characteristic features; 
\begin{enumerate}
\item The diagonal elements are zero due to antisymmetric properties $f_{ll'}=-f_{l'l}$.
\item There is no $CP$ violation because the three phases in the mass matrix 
are absorbed by the three Majorana neutrino fields. However, if we take into account of two 
loop effects there appear tiny non-zero masses at diagonal parts and it is possible to obtain 
$CP$ violation in this model\cite{Zee3}. 
\item If we assume naively the relations $f_{e\mu}\simeq f_{e\tau}\simeq f_{\mu\tau}$, 
it is expected that the relation 
$m_{\mu\tau}\simeq m_{e\tau}\gg m_{e\mu}$\cite{Smi} 
due to $m_{ij}\propto f_{ij}(m_i^2-m_j^2)$. 
However the mass and mixing matrices are  
 inconsistent with the result of CHOOZ's experiments\cite{CHOOZNew}. 
\end{enumerate} 
\par 
In this analysis we call the mass matrix given by Eq.(\ref{Mnu}) as 
Zee-type neutrino mass matrix and 
we investigate the consistent solution of Zee type mass matrix with experiments 
generally by relaxing the relation 
$f_{e\mu}\simeq f_{e\tau}\simeq f_{\mu\tau}$\cite{Fra}.
\section{Bi-maximal MNS Matrix}
The mass matrix (\ref{Mnu}) is symmetric and diagonalization is done by the 
MNS mixing matrix
\begin{equation}
U_{MNS}=
\left(\matrix{c_1c_3 & s_1c_3  & s_3 \cr
      -s_1c_2-c_1s_2s_3 & c_1c_2-s_1s_2s_3 & s_2c_3 \cr
       s_1s_2-c_1c_2s_3 & -c_1s_2-s_1c_2s_3 & c_2c_3 \cr}
\right) \; .
\label{U}
\end{equation}
We obtain the following conditions due to the property of Zee-type mass matrix;
\begin{eqnarray}
(1,1)\:\:& & ~ m_1c_1^2c_3^2+m_2s_1^2c_3^2+m_3s_3^2=0 \;, \nonumber\\ 
(2,2)\:\:& & ~ m_1(s_1c_2+c_1s_2s_3)^2+m_2(c_1c_2-s_1s_2s_3)^2+m_3s_2^2c_3^2=0
\;, \nonumber\\ 
(3,3)\:\:& & ~ m_1(s_1s_2-c_1c_2s_3)^2+m_2(c_1s_2+s_1c_2s_3)^2+m_3c_2^2c_3^2=0
\;,
\nonumber 
\end{eqnarray}
where $m_i(i=1,2,3)$ is the eigenvalues of Eq.(\ref{Mnu}).
Then the relations 
\begin{eqnarray}
m_2=-\frac{\cos^2\theta_1-\tan^2\theta_3}{\sin^2\theta_1-\tan^2\theta_3}m_1, & &\qquad 
m_3=-m_1-m_2 \;\nonumber\\
\cos2\theta_1\cos2\theta_2\cos2\theta_3&=&
\frac{1}{2}\sin2\theta_1\sin2\theta_2(3\cos^2\theta_3-2)\sin\theta_3
\label{condition}
\end{eqnarray}
are derived. 
It is noted that the hierarchical mass solution like $m_3\gg m_2 \gg m_1$ 
is prohibited in this model by the traceless condition. 
\par
Here we input the requisite condition $\theta_2\simeq \pi/4$ to obtain large 
mixing suggested by the atmospheric neutrino experiments\cite{Atm}. 
{}From Eq.(\ref{condition}) the possible solutions are 
$\theta_3\simeq 0$, $\theta_1\simeq 0$ or $\theta_3\simeq \arctan\sqrt{1/2}$. 
However the latter two cases except for $\theta_1\simeq\theta_3\simeq 0$ require
 $|U_{e3}|> 0.22$ and this solution is contradicting with the CHOOZ experiment\cite{CHOOZNew}.
Then it is enough to investigate the case of  $\theta_3\simeq 0$.
There are two hierarchical mass squared differences $\Delta m^2_{atm}\gg \Delta m_{solar}$ 
and in this model 
the unique solution is given for $\theta_3\simeq 0$ in Eq.(\ref{U}).
\par
The degenerate solutions with two masses are given for (1)$\theta_1\simeq 0$, (2) $\theta_1\simeq \arctan 1/\sqrt{2}$ and (3)$\theta_1 \simeq \pi/4$. 
The former two cases are inconsistent with the experimental data\cite{JMST}. 
The consistent solution is given only for the last case, 
which implies the solution $|m_1| \simeq |m_2| \gg |m_3| \quad , \:\:
\Delta m^2_{13}\simeq \Delta m^2_{23}=\Delta m^2_{atm}\ ,\quad 
\Delta m^2_{12}=\Delta m^2_{\odot}$.  The mixing matrix is obtained as 
\begin{equation}
U^{Zee}=\left(\matrix{\frac{1}{\sqrt{2}} & \frac{1}{\sqrt{2}} & 0 \cr
-\frac{1}{2} & \frac{1}{2} & \frac{1}{\sqrt{2}} \cr
\frac{1}{2} & -\frac{1}{2} &
\frac{1}{\sqrt{2}} \cr}\right) .
\end{equation}
This corresponds to the MSW or vacuum large angle solution in solar neutrino 
experiments.
Only the third case($\theta_1\simeq \pi/4,\theta_2\simeq\pi/4, \theta_3\simeq 0$) can  give 
a consistent solution and the MNS mixing matrix is bi-maximal. 
The neutrino masses are
$|m_1|\simeq |m_2|\simeq \sqrt{\Delta m_{atm}^2}, \quad 
|m_3|\simeq \frac{\Delta m^2_\odot}{2\sqrt{\Delta m_{atm}^2}}$
and an anti-hierarchical mass structure is obtained. 
The relation $ m_1\simeq -m_2$ implies pseudo-Dirac solution. 
For the parameters in the Zee model 
the consistent solution with experiments suggests a hierarchical relation as 
$f_{e\mu} \gg f_{e\tau} \gg f_{\mu\tau}$.
\section{Summary}
The unique solution of the model is the bi-maximal one, which corresponds to 
MSW or vacuum large mixing angle solution in solar neutrino oscillation.
The neutrino masses are 
$|m_1|\simeq |m_2|\simeq 0.02\sim 0.08 {\rm eV}(|m_3|\ll |m_1|\simeq |m_2|)$.
The neutrinoless double $\beta$ decay is induced by the (1,1) element in the 
mass matrix and  
$\langle m_\nu \rangle = | \sum_i U^2_{ei}m_{\nu i}|=|(1,1)| 
\approx 0$ prohibits the neutrinoless double $\beta$ decay. 
\par
The solution  implies the relation 
$f_{e\mu}\gg f_{e\tau} \gg f_{\mu\tau}$ in the Zee model.
The conservation of the new lepton number 
$L_{new}\equiv L_e-L_\mu-L_\tau$ requires 
$f_{\mu\tau}=0$ means $m_{\mu\tau}=0$. 
The realistic model is suggested in the extended Zee model with $L_{new}=-2$ doubly charged 
Higgs scalar field and 2-loop effects derive non-zero $m_{\mu\tau}$\cite{Josi}. 
Also the two loop effects in the original Zee model are studied\cite{Zee1}.
\par
Also the models which give the Zee type mass matrix are discussed in the viewpoint of $R$-parity violating SUSY model\cite{Cheung} or gauge mediated SUSY model\cite{Haba}. 
Also three active and one sterile neutrino model is discussed in Ref.\cite{Gaur}. 
The phenomenology of $h^{+}$ is also analyzed\cite{Ng}.
\par
The bi-maximal solution in Zee type mass matrix requires 
two degenerate heavy masses and one light neutrino. 
Solar neutrino oscillation experiment can resolve which mixing matrix is viable. 


\begin{thebibliography}{99}
\bibitem{Atm}
the Super-Kamiokande Collaboration, Talk given by K.Obayashi(ICRR) at this conference.
%
\bibitem{Sol}         
The Super-Kamiokande Collaboration, Talk given by Y.Fukuda(ICRR) at this conference.
%
\bibitem{JMST}
This talk was given by M.Matsuda based on the work given by 
C. Jarlskog, M. Matsuda, S. Skadhauge and M. Tanimoto, Phys. Lett {\bf B449}, 
(1999) 240. Please see this on the detailed analyses. 
%
\bibitem{Zee1}
A. Zee, Phys. Lett. {\bf 93B}, (1980) 389; {\bf 161B}, (1985) 141; 
Please also see the references cited in Ref.\cite{JMST} for the works on the Zee model.  
%
\bibitem{Zee3}
D. Chang and A. Zee, Phys.Rev. {\bf D61},(2000) 071303.
\bibitem{Smi}
A. Yu. Smirnov and Z. Tao, Nucl. Phys. {\bf B426}, (1994) 415;
A.Y.Smirnov and M.Tanimoto, Phys. Rev.  {\bf D55}, (1997) 1665.
\bibitem{CHOOZNew}  
The CHOOZ collaboration, Phys. Lett. {\bf B420}, (1998) 397.
%
\bibitem{Fra}
Similar analyses are given by P.H. Frampton and S.L.Glasow, Phys. Lett. {\bf B461}, (1999) 95.
%
\bibitem{Josi}
A.S.Josipura and S.D.Rindani,Phys. Lett. {B464},(1999) 239.
%
\bibitem{Cheung}
K.Cheung and Q.C.W.Kong, hep-ph/9912238.
\bibitem{Haba}
N.Haba, M.Matsuda and M.Tanimoto, Phys. Lett. {\bf 478}, (2000) 351.
%
\bibitem{Gaur}
N.Gaur, A.Ghosal, E.Ma, P.Roy, Phys. Rev. {\bf D58}, (1998) 071301.
%
\bibitem{Ng}
G.C.McLaughlin and J.N.Ng, Phys. Lett. {\bf B456}, (1999) 224.
%
\end{thebibliography}
\end{document}